\documentstyle[preprint,aps]{revtex}
\begin{document}
\tightenlines
\input{epsf}
\def\kk{${\rm K}^{0}-\overline{\rm K}^{0}$}
\def\bb{${\rm B}^{0}-\overline{\rm B}^{0}$}
\def\dd{${\rm D}^{0}-\overline{\rm D}^{0}$}
\def\bsg{B.R.($b \rightarrow s\gamma$) }
\def\ac{$\alpha_{\Xi}\,$}
\def\bc{$\beta_{\Xi }\,$}
\def\bcp{$\overline{\beta}_{\Xi}\,$}
\def\gc{$\gamma_{\Xi}\,$}
\def\gcp{$\overline{\gamma}_{\Xi}\,$}
\def\pc{$\phi_{\Xi}\,$}
\def\cascade{$\Xi\; $}
\def\lam{$\Lambda^{0\;}$}
\def\cm{$\Xi^{-}\;$}
\def\cp{$\overline{\Xi}^{+}\;$}
\def\pim{${\pi}^{-}\;$}
\def\ra{$\rightarrow\; $}
\def\omega{$\Omega\; $}
\def\om{$\Omega^{-}\;$}
\def\km{$K^{-}\;$}
\def\chisq{$\chi^{2}\;$}
\def\gev{GeV/$c\;$}

\preprint{LBL-1034, IIT-HEP-03/1}
\title {Measurement of Decay Parameters for $\Xi^{-} \rightarrow \Lambda
\pi^{-}$ Decay}

\author{
A.~Chakravorty,$^{4}$ H.T.~Diehl,$^{7, a}$ J.~Duryea,$^{6, b}$
G.~Guglielmo,$^{6,a}$ K.~Heller,$^6$ P.M.~Ho,$^{2, c}$
C.~James,$^3$ K.~Johns,$^{6, d}$ D.M~Kaplan,$^{4}$ M.J.~Longo,$^5$
K.B.~Luk,$^{1,2}$ R.~Rameika,$^3$ H.A.~Rubin,$^{4}$ S.~Teige,$^{7,
e}$ G.B.~Thomson,$^7$ Y.~Zou$^7$  (Fermilab E756 Collaboration)}

\address{$^1$ Department of Physics, University of California, Berkeley, CA 94720\\
$^2$ Physics Division, Lawrence Berkeley National Laboratory,
    University of California, Berkeley, CA 94720\\
$^3$ Fermilab, Batavia, IL 60510\\
$^4$ Physics Division, Illinois Institute of Technology, Chicago, IL 60616\\
$^5$ Department of Physics, University of Michigan, Ann Arbor, MI 48109\\
$^6$ School of Physics, University of Minnesota, Minneapolis, MN 55455\\
$^7$ Department of Physics and Astronomy, Rutgers--The State
University, Piscataway, NJ 08854}
\maketitle
\begin{abstract}
Based on 1.35 million polarized $\Xi^{-}$ events, we measure the
parameter $\phi_{\Xi}$ to be $-1.61^{\circ} \pm 2.66^{\circ} \pm
0.37^{\circ}$ for the $\Xi^{-} \rightarrow \Lambda \pi^{-}$ decay.
New results for the parameters $\beta_{\Xi}$ and $\gamma_{\Xi}$
are also presented. Assuming that the CP-violating phase-shift
difference is negligible, we deduce the strong phase-shift
difference between the P-wave and S-wave amplitudes of the
$\Lambda\pi$ final state to be $3.17^{\circ} \pm 5.28^{\circ} \pm
0.73^{\circ}$. This strong phase-shift difference reduces the
theoretical uncertainty in estimating the level of CP violation in
$\Xi$-hyperon decay.
\end{abstract}

\vspace*{0.2in} 


Breakdown of $CP$ invariance is well known in the weak decays of
the neutral $K$ meson \cite{cronin,ktev} and has been recently
established for the $B$ meson as well \cite{belle}. A deeper
understanding of this effect, which is evidence of a subtle
difference between the dynamics of particles and antiparticles, is
one of the central issues of current-day particle physics.
Complementary to $CP$ violation in the mesons is a similar effect
in the nonleptonic hyperon decays. This has been the subject of
theoretical discussion \cite{he,pakvasa} and ongoing experimental
searches \cite{luk,chauvat,hypercp}. Current models generally
predict the breakdown of $CP$ symmetry in strange-baryon decays
due to the different dynamics in the decay of a hyperon and its
antiparticle. In general, observing such a $CP$-odd effect
requires a strong phase-shift difference, which is the subject of
this paper.

For the $\Xi^{-}\rightarrow \Lambda \pi^{-}$ decay, the orbital
angular momentum of the $\Lambda \pi^{-}$ must be either $L$ = 0
\nobreak (S-wave) or  $L$ = 1 (P-wave) \cite{leeyang}. We write
the S- and P-wave amplitudes as \cite{he,wise}:
\begin{eqnarray}
S = |S|\exp[i(\delta_{S} + \eta_{S})],\,\, P =
|P|\exp[i(\delta_{P} + \eta_{P})]\,,
\end{eqnarray}
where $\delta_{S}$, $\delta_{P}$ are the strong rescattering
phases and $\eta_{S}$, $\eta_{P}$ are weak $CP$-violating phases.
The $CP$ asymmetry of the $\Xi$ decay \cite{he,pakvasa},
\begin{eqnarray}
A_{\Xi}= \frac{\alpha_{\Xi} +
{\alpha}_{\overline{\Xi}}}{\alpha_{\Xi} -
{\alpha}_{\overline{\Xi}}}\,,
\end{eqnarray}
with $\alpha_{\Xi}$ ($\alpha_{\overline{\Xi}}$) being a Lee-Yang
parameter of the hyperon (antihyperon) decay \cite{leeyang}, is
given by:
\begin{eqnarray} \label{eq:strong}
 A_{\Xi} \simeq - \tan(D) \sin(d_{CP}),
\end{eqnarray}
where $D = \delta_{P} - \delta_{S}$ and $d_{CP} = \eta_{P} -
\eta_{S}$ are phase-shift differences. Theoretical estimates of
$D$ vary between $-3^\circ$ and $16^\circ$ \cite{wise,nath}. The
model-dependent $CP$-violating phase shifts are generally
estimated to be a couple of orders of magnitude smaller than the
strong phase shifts \cite{pakvasa}. $D$ can thus be deduced from
the decay parameter $\beta_{\Xi}$
expressed as
\begin{eqnarray}
\label{eq:beta} \beta_{\Xi} = \alpha_{\Xi} \tan(D + d_{CP}) \simeq
\alpha_{\Xi} \tan(D).
\end{eqnarray}

The parameter $\beta_{\Xi}$ is conveniently determined in terms of
a phase $\phi_\Xi ={\rm \tan}^{-1}(\beta_\Xi/\gamma_\Xi)$, where
$\gamma_\Xi$ is another decay parameter of the decay such that
$\alpha_{\Xi}^{2} + \beta_{\Xi}^{2} + \gamma_{\Xi}^2$ = 1.
The world average is $\phi_\Xi = 4^\circ \pm 4^\circ$, taken over
several measurements with the largest sample using 20,865 events
\cite{bensinger}. In this Letter we report a new measurement of
$\phi_{\Xi}$ from a data set of 1.35 million polarized $\Xi^{-}
\rightarrow \Lambda \pi^{-}$ decays.

Our data come from Experiment 756
at Fermi National Accelerator Laboratory
 \cite{ho,diehl,duryea,duryea1}. Polarized
$\Xi^{-}$ hyperons with transverse momenta $p_{T}$ spanning 0.5 to
1.5 GeV/$c$ and momentum fraction $x_{F}$ from 0.3 to 0.7 were
produced by the collision of unpolarized 800-GeV/$c$ protons
incident upon a beryllium target at an angle with respect to the
vertical axis. The secondary beam was momentum- and sign-selected
by a curved collimator inside a dipole magnet. The current in the
magnet was set to yield a field integral ${\int}B dl$ of $15.30
\pm 0.15$ T-m, as measured with a Hall probe, during collection of
the data used in the present analysis.

The momenta of the proton and the pions from the decay sequence
$\Xi^- \rightarrow \Lambda \pi^-, \Lambda \rightarrow p \pi^-$
were measured with eight planes of silicon strip detectors,
nine multiwire proportional chambers, and two dipole analyzing
magnets that deflected charged particles in the horizontal plane.
The polarity of these magnets could be reversed by switching the
direction of the applied current. The \cm triggers and veto
scintillation counters have been described in  \cite{luk}. Data
were taken for two sets of production angles averaging $+2.4$ and
$-2.4$ mrad, respectively, and the analyzing magnet currents were
set to $\pm$2500A. The $\Xi^{-}$ polarization and magnetic moment
from these data have been reported \cite{duryea,duryea1}.

The events were analyzed off-line using a reconstruction program
that determined tracks and kinematic variables from the chamber
hits. Events were required to satisfy the three-track, two-vertex
topology corresponding to a $\Xi^{-}$ decay-sequence hypothesis.
The geometric $\chi^{2}$ for the topological fit was required to
be less than 100 for a mean of 30 degrees of freedom. The proton
and pion from $\Lambda$ decay were required to have a $p\pi$
invariant mass within 3.5 standard deviations (8 MeV/$c^2$) of the
$\Lambda$-decay peak at 1.116 GeV/$c^2$, and the $\Lambda \pi$
invariant mass was required to be within 5 standard deviations (14
MeV/$c^2$) of the $\Xi^{-}$-decay peak at 1.321 GeV/$c^2$. The
momenta of the reconstructed $\Xi^{-}$ candidates were required to
be between 240 and 450 GeV/$c$.
The $\Xi^{-}$ momentum was required to trace back to within 0.63
cm of the center of the beryllium target. The $\Lambda$-decay
vertex was required to be downstream of the $\Xi^{-}$-decay
vertex, and both vertices were required to be in a fiducial region
0.25 m downstream of the collimator exit and 0.31 m upstream of
the multiplicity counter located at 23.31 m from the collimator
exit.

The decay parameters are related to the experimental observables
through the  polarization and angular distribution of the daughter
baryon. For $\Xi^{-} \rightarrow \Lambda \pi^{-}$, we have
\cite{leeyang}:
\begin{eqnarray}
\label{eq:polrel}\vec{P}_{\Lambda} = \frac{(\alpha_{\Xi}\! +
\vec{P}_{\Xi}\cdot\hat{\Lambda})\hat{\Lambda}\! +
\beta_{\Xi}(\vec{P}_{\Xi}\!\!\times\hat{\Lambda}) \! +
\gamma_{\Xi}\hat{\Lambda}\!\!\times(\vec{P}_{\Xi}\!\!\times\hat{\Lambda})}{1
+ \alpha_{\Xi}\vec{P}_{\Xi}\cdot\hat{\Lambda}}\,,
\end{eqnarray}
where $\vec{P}_{\Lambda}$ is the polarization of the $\Lambda$
hyperon in its rest frame, $\hat{\Lambda}$ is the momentum
direction of the $\Lambda$ in the \cm rest frame, and
$\vec{P}_{\Xi}$ is the \cm polarization in the \cm frame. The
helicity-frame axes, specified event-by-event in the \cm rest
frame, are defined as
\begin{eqnarray}
\label{eq:helicityframe}
\hat{X}=\frac{\vec{P}_{\Xi}\!\!\times\hat{\Lambda}}{|\vec{P}_{\Xi}\!\!\times\hat{\Lambda}|},
\;\;\hat{Z}= \hat{\Lambda},  \; \; \hat{Y}=
\hat{Z}\!\!\times\hat{X}\,.
\end{eqnarray}
Then, for the $\Lambda \rightarrow p \pi^{-}$ decay,  the angular
distribution of the proton in the $\Lambda$ rest frame can be
projected onto the helicity-frame axes to give \cite{cool}
\begin{eqnarray}
\label{eq:a-distrib} \frac{dn}{d \,\cos\theta_{pZ}}= \frac{1}{2}(1
+
 \alpha_{\Lambda}\alpha_{\Xi}\cos\theta_{pZ})\,,
\end{eqnarray}
\begin{eqnarray}
\label{eq:b-distrib} \frac{dn}{d \,\cos\theta_{pX}}= \frac{1}{2}(1
+ \frac{\pi}{4}
 \alpha_{\Lambda}\beta_{\Xi}P_{\Xi}\cos\theta_{pX})\,,
\end{eqnarray}
\begin{eqnarray}
\label{eq:g-distrib}  \frac{dn}{d \,\cos\theta_{pY}}=
\frac{1}{2}(1 +\frac{\pi}{4}
 \alpha_{\Lambda}\gamma_{\Xi}P_{\Xi}\cos\theta_{pY})\,,
\end{eqnarray}
where $\theta_{pi}$ is the angle between the proton momentum in
the $\Lambda$ rest frame and the $i$th helicity-frame axis.
The ratio of the slopes of the $\cos\theta_{pX}$ and
$\cos\theta_{pY}$ distributions provides a measurement of $\tan
\phi_{\Xi}= \beta_\Xi / \gamma_{\Xi}$.

Conservation of parity in the strong interactions dictates that
any \cm polarization be normal to the production plane at the
target. The precession angle $\Phi$ of this polarization relative
to the \cm momentum at the collimator exit, after the \cm hyperons
pass through the vertically directed magnetic field in the
collimator, is given by
\begin{eqnarray}
\Phi = \frac{q}{\beta m_{\Xi} c^2} \left\{ -\frac{\mu_{\Xi}
m_{\Xi}}{\mu_{N} m_{p}} - 1\right\} \int{\! Bdl}\,\,,
\end{eqnarray} where $q$, $\mu_{\Xi}$ and $m_{\Xi}$ are the electric
charge, magnetic moment and mass of the $\Xi^{-}$ respectively,
$\beta \approx$ 1, $\mu_N$ is the nuclear magneton, and $m_p$ is
the proton mass. As reported in \cite{duryea}, the polarization of
$\Xi^{-}$ for this data set was confined in the horizontal plane
and approximately 10$\%$ in magnitude. We used $\mu_{\Xi}= -0.6505
\pm 0.0025\ {\mu_{N}}$ \cite{PDG} to determine $\Phi =
-16.68^\circ \pm 0.86^{\circ}$ and hence the orientation of the
$\Xi^-$ polarization in the spectrometer. The helicity-frame axes
defined in Eq.~\ref{eq:helicityframe} were calculated from the
reconstructed $\Lambda$ momentum.

The slopes of the $\cos\theta_{pX}$ and $\cos\theta_{pY}$
distributions in Eqs.~\ref{eq:b-distrib} and \ref{eq:g-distrib}
were measured using the Hybrid Monte Carlo (HMC) method
\cite{bunce} which factored out the acceptances. Up to 200 HMC
events were generated per real event for each distribution. These
HMC events were uniformly distributed in $\cos\theta_{pX}$ (${\rm
\cos}\theta_{pY}$), with the rest of the kinematic quantities such
as decay vertices and momentum of the $\Lambda$ taken from the
associated real event. The event was included in the measurement
when ten of its associated HMC events satisfied all of the
requirements used to simulate the triggers, the geometry and
inefficiencies of the spectrometer. A $\chi^2$ minimization based
on the comparison of weighted HMC data with real events was used
to determine the slope of each $\cos \theta$ distribution.

\begin{figure}[hbt]
\vspace{-0.1cm}
\vspace*{-1cm} \indent a) \centerline{\epsfxsize=4in
\epsfbox{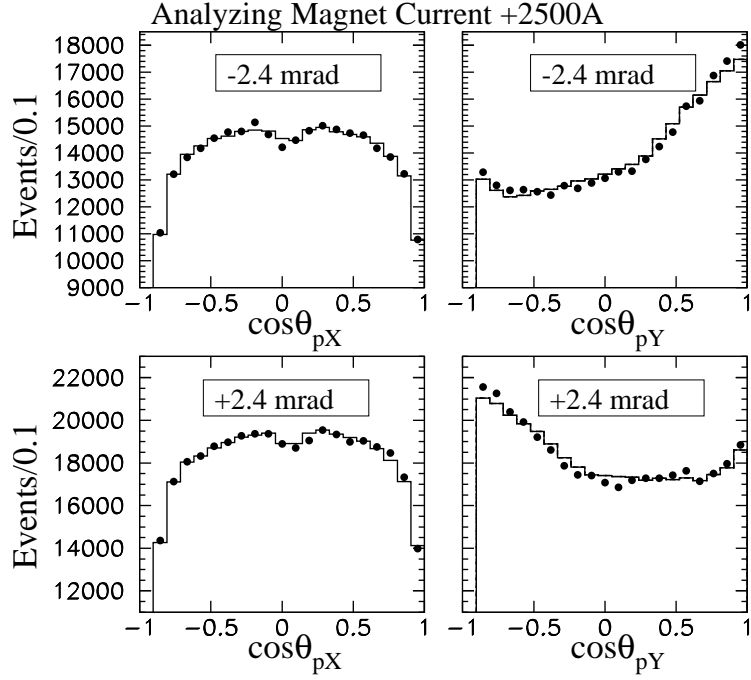}} \vspace*{-1.3cm} b) \centerline{\epsfxsize=4in
\epsfbox{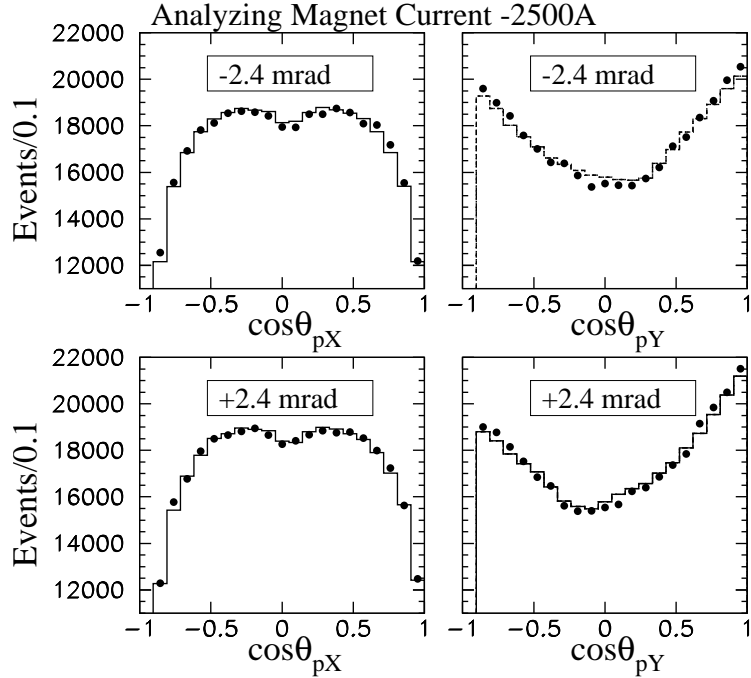}} \vspace{-0.5cm} \caption{$\cos
\theta_{pX}$~and~$\cos \theta_{pY}$~data~distributions~(points)
with corresponding weighted HMC events (histograms) for a) the
+2500A and b) the -2500A current setting of the analyzing magnets.
} \label{fig:2500A}
\end{figure}

The data were analyzed separately as four streams: data taken at
positive and negative production angles for each polarity of the
analyzing magnets. The measured slope for a given distribution is
comprised of the true slope plus a bias term \cite{ho}. The bias,
resulting from imperfections in the experimental apparatus and
reconstruction effects not fully simulated in the analysis, is not
sensitive to the reversal of the production angle. The
polarization, however, changes sign when the production angle is
reversed. The helicity-frame axes in Eq.~\ref{eq:helicityframe}
transform as $\hat{X} \rightarrow -\hat{X}$, $\hat{Y} \rightarrow
-\hat{Y}$, and $\hat{Z} \rightarrow \hat{Z}$ when the polarization
changes sign. A bias term in the $X$-direction ($B_{X}$) will thus
reverse sign in the transformed helicity frame. If the slope
measured for the positive (negative) production angle is $S_{+}$
($S_{-}$), the true slope and bias are extracted using
\begin{eqnarray}
\frac{\pi}{4} \alpha_{\Lambda}\beta_{\Xi}P_{\Xi} = \frac{S_{+} +
S_{-}}{2}, \, \, B_{X} = \frac{S_{+} - S_{-}}{2}.
\end{eqnarray}
Similar expressions can be written for
$\alpha_{\Lambda}\gamma_{\Xi}P_{\Xi}$ and the $Y$ bias ($B_{Y}$).
Fig.~\ref{fig:2500A} shows the $\cos\theta_{pX}$ and
$\cos\theta_{pY}$ distributions for each stream and the
corresponding weighted HMC fit. The HMC analysis was repeated with
26 different random number seeds and the results were averaged.
Table~\ref{tab:asymmetry} shows the extracted slopes and bias
terms of the $\cos\theta_{pX}$ and $\cos\theta_{pY}$ distributions
for each polarity of the analyzing magnet.

To show that we were able to determine the correct axes, we
measured the slope $\alpha_{\Lambda}\alpha_{\Xi}$ of the
distribution in Eq.~\ref{eq:a-distrib}. Our result, which does not
include systematic studies, was $\alpha_{\Lambda}\alpha_{\Xi} =
-0.305 \pm 0.002$, consistent with the world average \cite{PDG}.
As a further check, we measured the magnitude of the $\Xi^{-}$
polarization by projecting the angular distribution of the proton
onto the laboratory axes and summing the components (a technique
discussed in \cite{ho,duryea}).  We then calculated
$\frac{\pi}{4}\alpha_{\Lambda}\gamma_{\Xi}P_{\Xi} = 0.043 \pm
0.003$, in agreement with the results in Table \ref{tab:asymmetry}
and in \cite{duryea}.

By taking the ratio of the measured values of $\frac{\pi}{4}
\alpha_{\Lambda}\beta_{\Xi}P_{\Xi}$ and $\frac{\pi}{4}
\alpha_{\Lambda}\gamma_{\Xi}P_{\Xi}$, the phase $\phi_{\Xi}$ was
determined to be ${-1.28}^\circ \pm {3.86}^\circ$ for the +2500A
data and ${-1.93}^\circ \pm {3.68}^\circ$ for the -2500A data. The
good agreement for two settings of the analyzing magnets provides
a systematic check of the analysis method.

\begin{figure}[hbt]
\centerline{\epsfxsize=4.5in \epsfbox{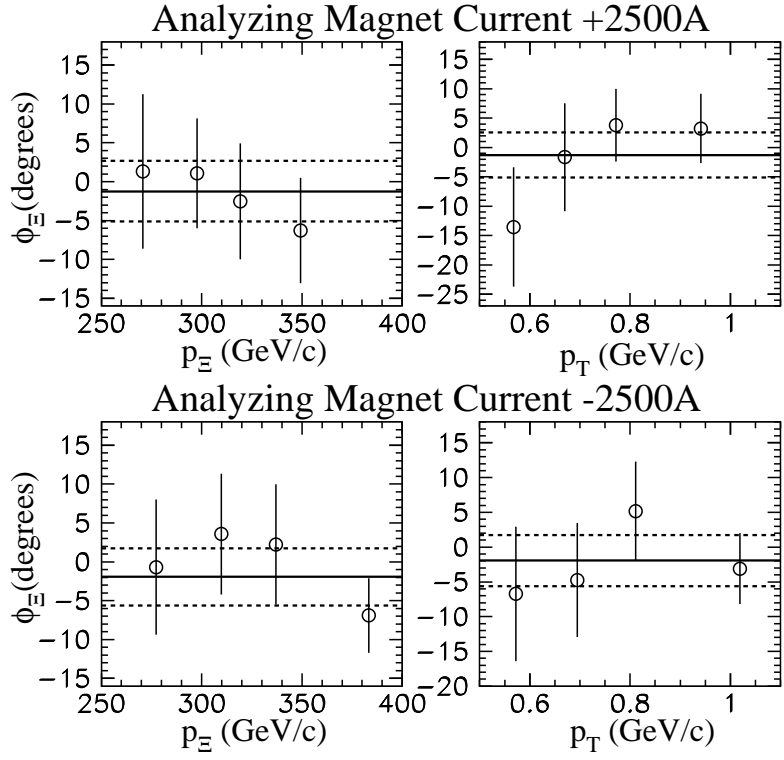}}
\caption[$ \phi_{\Xi}$  as a function of $P_{T}$]{$\phi_{\Xi}$  as
a function of $p_\Xi$ (left) and $p_{T}$(right) for +2500A (top)
and -2500A (bottom) settings of the analyzing magnets. Solid and
dotted lines indicate the central measurement and its statistical
error.} \label{ptbsig}
\end{figure}

The measured $\phi_{\Xi}$ displays no significant dependence on
either the $\Xi^{-}$ momentum, $p_{\Xi}$, or the transverse
momentum, $p_T$ (Fig.~\ref{ptbsig}).
A study  of the biases $B_{X}$ and $B_{Y}$ as a function of the
$\Xi^{-}$ momentum showed no significant dependence
\cite{chakravorty}. Variation of some of the data-selection
criteria allowed us to study systematic effects due to background,
poorly measured events, goodness of the track-fitting, and
resolution at the target. The systematic effect of the random
number seed in the HMC program was investigated.  In addition, we
studied the effect of changing the precession angle on the
measured $\phi_{\Xi}$ by varying the value of $\Phi$ by one
standard deviation. Table \ref{tab:systematics} summarizes our
estimates of the systematic uncertainties.

By averaging the measurements for the two settings of the
analyzing magnet, and adding the systematic errors in quadrature,
we obtain our final result:
\begin{eqnarray}
\nobreak \phi_{{\Xi}} = -1.61^\circ \pm 2.66^\circ \pm 0.37^\circ,
\nonumber
\end{eqnarray}
where the first error is statistical and the second systematic.
Our measurement is consistent with the current world average of
$4^\circ \pm 4^\circ$~\cite{PDG} and with zero. Using
\begin{eqnarray}
\beta_{\Xi} & = & (1 - \alpha^{2}_{\Xi})^{\frac{1}{2}}\sin{\phi_\Xi}, \\
\gamma_{\Xi} & = & (1 -\alpha^{2}_{\Xi})^{\frac{1}{2}}
\cos{\phi_\Xi},
\end{eqnarray}
where $\alpha_{\Xi }= -0.458\pm 0.012$ \cite{PDG}, we calculate
\begin{eqnarray}
\beta_{{\Xi}} & = & -0.025 \pm 0.042 \pm 0.006, \nonumber \\
\gamma_{{\Xi}} & = & +0.889 \pm 0.001 \pm 0.007.  \nonumber
\vspace*{-0.07in}
\end{eqnarray}

Both statistical and systematic errors on $\beta_{\Xi}$ are
dominated by the uncertainties in our measurement of $\phi_\Xi$,
whereas the systematic error on $\gamma_{\Xi}$ is due to the
uncertainty in $\alpha_\Xi$.

Using Eq.~\ref{eq:beta}, the phase-shift difference in the \cm
decay process is also deduced:
\begin{eqnarray}
D + d_{CP} = 3.17^{\circ} \pm 5.28^{\circ} \pm 0.73^{\circ}.
\nonumber
\end{eqnarray}
This measurement is consistent with zero and indicates that the
strong phase-shift difference of the final state in $\Xi^{-}
\rightarrow \Lambda \pi^{-}$ decay is small, in agreement with
recent calculations \cite{wise} but disagreeing with that of
\cite{nath}.

In conclusion, we have measured the parameter $\phi_{\Xi}$ using
1.35 million $\Xi^{-} \rightarrow \Lambda \pi^{-}$ decays. Our
result has a precision that is $\approx$ 1.5 times better than the
world average and $\approx$ 3.3 times better than the previous
best single measurement \cite{baltay}. With this result, we obtain
new values for $\beta_{\Xi}$ and $\gamma_{\Xi}$. The strong
phase-shift difference deduced from this measurement is consistent
with zero and thus imposes a limit on the level of $CP$ violation
in $\Xi^{-} \rightarrow \Lambda \pi^{-}$ hyperon decay.

We would like to acknowledge support by the U.S. Department of
Energy. K.B.L. was also partially supported by the Miller
Institute. The excellent assistance of the Fermilab staff was
essential for the completion of the experiment.

\begin{table}[hbt]
\begin{center}
\caption{Extracted slopes and biases}\label{tab:asymmetry}
\begin{tabular}{ccccc}
\hline\hline
Anal.         & No. of & Mean $\Xi$  &    & \\
Magnet        & Events & Momen-      &    & \\
Current       &        & tum         &    & \\
(A)           &        &(GeV/c)      & $\frac{\pi}{4}
\alpha_{\Lambda}\beta_{\Xi}P_{\Xi}$ & $B_{X}$ \\ \hline
$+$2500 & 646774 & 309 & $-$0.0009$\pm$0.0028 & $+$0.0042$\pm$0.0028 \\
$-$2500 & 696337 & 326 & $-$0.0014$\pm$0.0026 & $-$0.0015$\pm$0.0026 \\
\hline
& & & $\frac{\pi}{4} \alpha_{\Lambda}\gamma_{\Xi}P_{\Xi}$ & $B_{Y}$\\
\hline
$+$2500 & 647029 & 309 & $+$0.0410$\pm$0.0024 &$+$0.0037$\pm$0.0024 \\
$-$2500 & 696633 & 326 &$+$0.0400$\pm$0.0022 & $+$0.0012$\pm$0.0022 \\
\hline\hline
\end{tabular}
\end{center}
\end{table}

\begin{table}[hbt]
\begin{center}
\caption{Systematic errors for the measurement of \pc}
\label{tab:systematics}
\begin{tabular}{cc}
\hline\hline
                          & Estimated Maximum\\
Study                & Systematic Error \\ \hline
Momentum Dependence & $< 0.1^\circ$ \\
$p_{T}$ Dependence  & $< 0.1^\circ$ \\
Selection Criteria  & $0.25^\circ$ \\
Random number seed  & $0.25^{\circ}$ \\
Precession Angle    & $0.11^{\circ}$ \\\hline Total           &
$0.37^\circ$ \\
\hline\hline
\end{tabular}
\end{center}
\end{table}

\begin{thebibliography}{30}

\bibitem[a]{fnal}Present Address: Fermilab, Batavia, IL 60510.

\bibitem[b]{harvard}Present Address: Brigham and Women's Hospital, Harvard
Medical School,  Boston, MA 02115.

\bibitem[c]{pmho}Present Address: Visa International, 3055 Clearview Way,
MS 3A, San Mateo, CA 94402.

\bibitem[d]{az}Present Address: Department of Physics, University of Arizona,
Tucson, AZ 85721.

\bibitem[e]{ind}Present Address: Department of Physics, Indiana University,
Bloomington, IN 47405.

\bibitem{cronin}
J.H. Christenson  {\it et al.}, Phys.\ Rev.\ Lett. {\bf 13}, 138
(1964).

\bibitem{ktev}
A. Alavi-Harati {\it et al.}, Phys.\ Rev.\ Lett. {\bf 83}, 2128
(1999); V. Fanti {\it et al.}, Phys.\ Lett.\ B {\bf 465}, 335
(1999).

\bibitem{belle}
A. Abashian {\it et al.}, Phys.\ Rev.\ Lett. {\bf 86}, 2509
(2001); B. Aubert {\it et al.}, Phys.\ Rev.\ Lett. {\bf 86}, 2515
(2001).

\bibitem{he}
X.-G. He {\it et al.}, Phys.\ Rev.\ D {\bf 61}, 071701 (2000);
J.F. Donoghue, X.-G. He, S. Pakvasa, Phys.\ Rev.\ D {\bf 34}, 833
(1986).

\bibitem{pakvasa}
S. Pakvasa, preprint hep-ph/9910232 (1999); J. Tandean and G.
Valencia, preprint hep-ph/0211165 (2002).

\bibitem{luk}
K.B. Luk {\it et al.}, Phys.\ Rev.\ Lett. {\bf 85}, 4860 (2000).

\bibitem{chauvat}
P. Chauvat {\it et al.}, Phys.\ Lett.\ B {\bf 163}, 273 (1985);
M.H. Tixier {\it et al.}, Phys.\ Lett.\ {\bf 212}, 523 (1988);
P.D. Barnes {\it et al.}, Phys.\ Rev.\ C {\bf 54}, 1877 (1996).

\bibitem{hypercp}
C. Dukes {\it et al.}, Nucl.\ Phys.\ Proc.\ Suppl. {\bf 75B}, 281
(1999).

\bibitem{leeyang}
T.D. Lee and C.N. Yang, Phys.\ Rev. {\bf 108}, 1645 (1957).

\bibitem{wise}
M. Lu, M.B. Wise, M.J. Savage, Phys.\ Lett.\ B {\bf 337}, 134
(1994); A. Datta and S. Pakvasa, Phys.\ Lett.\ B {\bf 344}, 430
(1995); A.N. Kamal, Phys.\ Rev.\ D {\bf 58}, 077501 (1998);
J.Tandean {\it et al.}, Phys.\ Rev.\ D {\bf 64}, 014005 (2001);
U.G. Meissner and J.A. Oller, Phys.\ Rev.\ D {\bf 64}, 014006
(2001).

\bibitem{nath}
R. Nath and A. Kumar, Nuovo\ Cimento {\bf 36}, 669 (1965).

\bibitem{bensinger}
J.R. Bensinger {\it et al.}, Nucl.\ Phys.\ B {\bf 252}, 561
(1985).

\bibitem{ho}
P.M. Ho {\it et al.}, Phys.\ Rev.\ D {\bf 44}, 3402 (1991); P.M.
Ho, Ph.D. Thesis, University of Michigan (1991).

\bibitem{diehl}
H.T. Diehl {\it et al.}, Phys.\ Rev.\ Lett. {\bf67}, 804 (1991);
H.T. Diehl, Ph.D. Thesis, Rutgers University (1990).

\bibitem{duryea}
J. Duryea {\it et al.}, Phys.\ Rev.\ Lett. {\bf 67}, 1193 (1991);
J. Duryea, Ph.D. Thesis, University of Minnesota (1990).

\bibitem{duryea1}
J. Duryea {\it et al.}, Phys.\ Rev.\ Lett. {\bf 68}, 768 (1992).

\bibitem{cool}
R.L. Cool {\it et al.}, Phys.\ Rev.\ D {\bf 10}, 792 (1974).

\bibitem{PDG}
Particle Data Group, K.\ Hagiwara {\it et al.}, Phys.\ Rev.\ D
{\bf 66}, 010001 (2002).

\bibitem{bunce}
G. Bunce, Nuc.\ Instr.\ and Meth.\ {\bf 172}, 553 (1980).

\bibitem{chakravorty}
A. Chakravorty, Ph.D. Thesis, Illinois Institute of Technology
(2000).

\bibitem{baltay}
C. Baltay {\it et al.}, Phys.\ Rev.\ D {\bf 9}, 49 (1974).


\end{thebibliography}
\end{document}